\documentclass{aastex}
\shorttitle{White Dwarfs in Polars}

\begin{document}
\title{The Low State Temperature Distribution and First Chemical Abundances
of White Dwarfs in Polars}

\author{Edward M. Sion }
\affil{Department of Astronomy and Astrophysics,
Villanova University,
Villanova, PA 19085,
USA}
\email{edward.sion@villanova.edu}

\begin{abstract}

During the low optical brightness states of AM Herculis systems (polars)
when accretion has declined to a very low value, the underlying magnetic
white dwarf photosphere can be modelled without the complication of
thermal bremstrahlung and cyclotron emission from the luminous accretion
column. The far ultraviolet spectra can be modelled with high gravity
solar composition photospheres. In this way, I present new temperatures
and the first chemical abundance estimates for the white dwarfs in three
selected polars from the IUE NEWSIPS archive. For the white dwarf in V834
Cen with T$_{eff} = 16,000$K, Si/H = 0.1 solar, C/H = 0.5 solar, for BY
Cam, T$_{eff} = 17,000$K, Si/H = 0.1, C/H = 5 solar and for RX J1313-32,
T$_{eff} = 22,000$K, Si/H = 0.1 solar, C/H = 0.1 solar. The temperature
distribution of 24 white dwarfs in polars with known temperatures above
and below the period gap is compared with the distribution of the white
dwarf temperatures in dwarf novae during quiescence. In both cases, the
magnetic white dwarfs in polars are significantly cooler than the
non-magnetics. For all CV white dwarfs, magnetic and non-magnetic with
T$_{eff} < 12,500$K, 91\% of the objects (10 out of 11) are magnetics in
polars. This suggests that long term accretion heating and cooling of
white dwarfs in polars differs from the effects of long term accretion in
non-magnetic disk accretors.

\end{abstract}
\keywords{Polars, White Dwarfs, Accretion}

\section{Introduction}

The physics of magnetic accretion at the highly magnetic white dwarf
surface in polars is poorly known. We do not yet understand the lateral
spread (sideways diffusion) and downward diffusion of accreted matter, and
the long term heating of the magnetic white dwarf (compared with the
heating due to disk accretion).  Yet it is these very systems which offer
us tantalizing possibilities for increasing our understanding through
studies during their low optical brightness states. This is because, for
reasons which remain unexplained, accretion declines to an extremely low
rate during these low states and hence the emission contribution from
thermal bremstrahlung and cyclotron sources from the accretion column is
no longer evident in their far ultraviolet spectra. All that is left to
observe is the bare magnetic white dwarf with relatively rapidly cooling
polar accretion regions. By using white dwarf model atmospheres to analyze
far UV spectra obtained with IUE and HST during low states, new insights
are possible on all of the above questions.

In this talk, I present the first results on the chemical abundances of
the accreted atmospheres of the white dwarfs in polars starting with the
prototype AM Herculis itself, V834 Cen, BY Cam and RX J1313-32. For
comparison, the metal abundance derived by Sion et al. (1998) for the
magnetic white dwarf in the pre-CV (pre-IP?) V471 Tauri will also be
compared. Next, I present the latest information on the low state
temperature distribution of the magnetic white dwarfs in polars and
compare their temperature distribution with the white dwarf temperatures
in dwarf novae. I do this in two ways: number versus T$_{eff}$
distribution histograms and by using the distribution function of magnetic
and non-magnetic white dwarf temperatures versus orbital period. 

\section{Chemical Abundances}

Our knowledge of chemical abundances derives from fitting solar
composition high gravity model atmospheres to the low state spectra of
polars. and varying the abundances of individual elements until the best
fit is achieved. The first abundances derived in this way were obtained by
DePasquale and Sion (2001) who carried out fits to the white dwarf in AM
Herculis during its low states at different times following the last
previous high states. On average, they found that T$_{eff} = 20,400$K at
UV minimum (when the field is perpendicular to the line of sight) and
23,500K at UV maximum (when the magnetic pole is facing the observer).
These temperatures are close to the values found by G"ansicke (1997).
DePasquale and Sion (2001) carried out abundance fits and found silicon
abundances, Si/H, from 14 spectra ranging between 0.001 x solar and 0.05 x
solar. There was no apparent difference in abundance with increasing time
since the high state nor between spectra at UV maximum and UV minimum. For
comparison, Sion et al. (1998) found that Si/H = 0.001 from fitting a
detected photospheric Si III (1206) absorption feature in the magnetic
white dwarf in the pre-CV V471 Tauri when the large accretion cap is
facing the observer (X-ray rotational phase 0.5).

Since polars are very faint during their low states, IUE observations were
possible for only a few systems. The archival spectra are quite noisy. A
search of the archive revealed three systems with spectra of marginal
quality for a photosphere analysis, V834 Cen, BY Cam and RX J1313-32. The
low state temperatures of these systems were first derived by G\"{a}nsicke
(1997). Here, I present new temperatures and estimate the chemical
abundances.  For RX J1313-32, the best-fitting model revealed T$_{eff} =
18,000$K $\pm1000$K, Si/H = 0.1 times solar, C/H = 0.1 times solar. This
model fit is displayed in figure 1. Sinilarly, for V834 Cen, the
best-fitting model yielded T$_{eff} = 16,000$K $\pm1000$K, log $g = 8$,
Si/H = 0.1 x solar and C/H = 0.5 times solar.  For BY Cam, the same
fitting procedure led to a best-fitting model with T$_{eff} =
22,000$K$\pm1000$K, log $g = 8$, Si/H = 0.1 x solar and C/H = 5 times
solar.

\section{The Low State Temperature Distribution of White Dwarfs in Polars}

Is there a difference in long term heating between the accreting white
dwarfs in polars and the accreting white dwarfs in disk systems? If so,
does it arise from a difference between a magnetic accretion geometry and
disk (tangential) accretion geometry? It was first noted by Sion (1991)
that magnetic white dwarfs in polars tended to be cooler, at the same
orbital period, than the disk-accreting white dwarfs in dwarf novae and
nova-like variables. This conclusion was bolstered by a re-examination of
an increased sample of white dwarf temperatures in polars presented in the
review by Sion (1999). Here, I discuss a sample size of magnetic accretors
one-third larger than in Sion (1999), including 3 new cool magnetic
degenerates in polars presented at this meeting (see the papers by Schmidt
2003, Szkody 2003, Ferrario et al. 2003, this volume), two of which,
SDSS1553+5516 and SDSS13254+0320, were recently discovered in the Sloan
Digital Sky Survey (Szkody et al. 2003).

For systems below a period of 2 hours (i.e., below the period gap) the
average temperature of magnetic white dwarfs in polars is $<T_{eff}> =
13,710$K while non-magnetic white dwarfs in dwarf novae have $<T_{eff>} =
15,547$K. For the magnetic white dwarfs in polars above a period of 3
hours (i.e., above the period gap), $<T_{eff>} = 17350$K while for
non-magnetic white dwarfs in dwarf novae, $<T_{eff}> = 31,182$K.
In figure 2, I display the temperatures (black triangles) of all of the
magnetic white dwarfs in polars versus their orbital period in minutes.
In figure 3, I display the temperatures of all of the white dwarfs in
dwarf novae during quiescence (filled circles) together with the magnetic
white dwarfs in polars (filled triangles)

In fig.4, I display a histogram of number versus T$_{eff}$ for the
magnetic white dwarfs in polars during low states (solid lines) in
comparison with the white dwarfs in disk-accreting dwarf novae during
quiescence (dashed lines). Note that below 12,500K, 98\% of the white
dwarfs are in polars. Note that for 11 magnetic and non-magnetic sytems
with white dwarf T$_{eff} < 12,500$K, 91\% of the objects are magnetic WDs
in polars. While we cannot dismiss a possible role for observational
selection, we believe the evidence is compelling enough to support our
conclusion that magnetic white dwarfs in polars are cooler (and may cool
differently) than disk-accreting, non-magnetic white dwarfs in dwarf
novae. The effect of long term accretion heating on magnetic white dwarfs
in polars may be different than the effect of long term accretion heating
in disk accretors.

\noindent Figure Captions

\noindent Fig. 1 - The observed IUE low state spectrum (SWP56879) of RX
J1313-32 (flux F$_{\lambda}$ versus wavelength (\AA)) with the
best-fitting white dwarf model atmosphere (thicker black line) having
T$_{eff} = 18,000$K, log $g = 8$, with Si/H = 0.1 times solar, and C/H =
0.1 times solar.

\noindent Fig.2 - The orbital period (minutes) versus white dwarf
T$_{eff}$ (K) diagram 
for the white dwarfs in polars (triangles) with temperatures determined
during their low states.

\noindent Fig.3 - The orbital period (minutes) versus white dwarf
T$_{eff}$ (K) diagram 
for the white dwarfs in dwarf novae during quiescence (filled circles) and
the white dwarfs in polars during low states
(triangles).

\noindent Fig. 4 - A histogram of number versus T$_{eff}$ for the magnetic
white dwarfs
in polars during low states (solid lines) in comparison with the white
dwarfs in disk-accreting dwarf novae during quiescence (dashed lines).
Note that below 12,500K, 98\% of the white dwarfs are in polars.

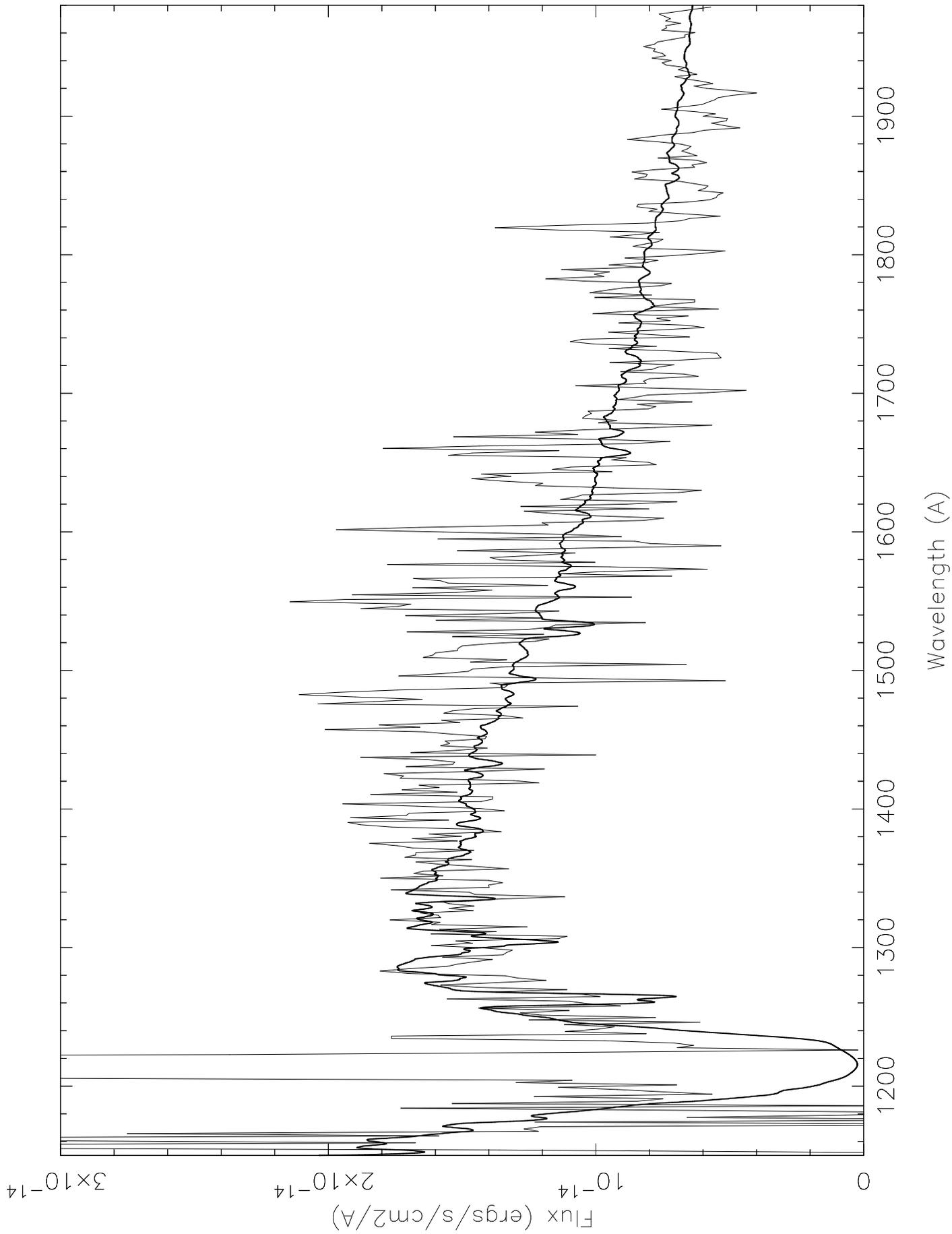
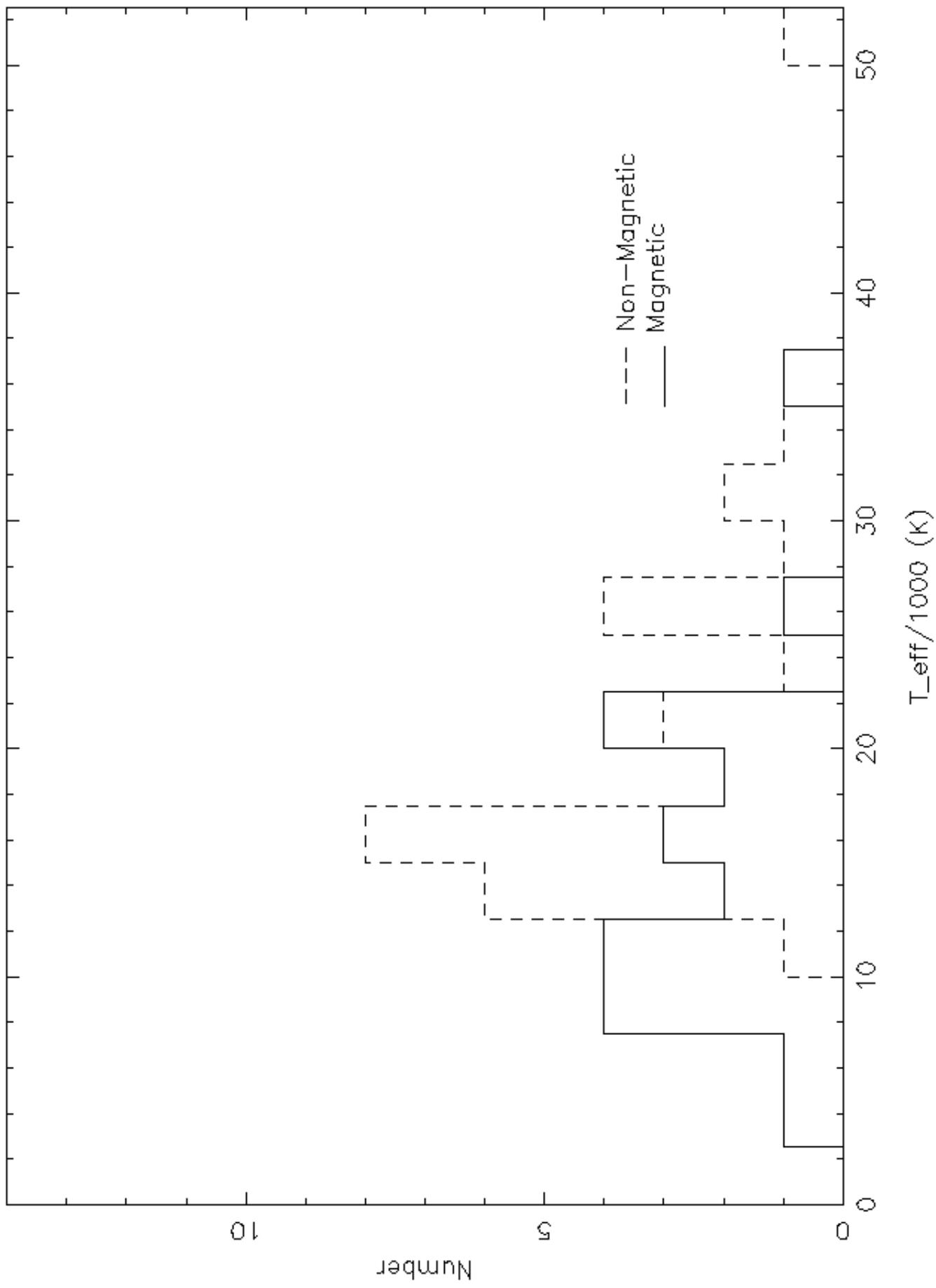
\end{document}